# Establishing an Online Access Panel for Interactive Information Retrieval Research

Dagmar Kern
GESIS – Leibniz Institute for the Social Sciences
Unter Sachsenhausen 6-8
50667 Cologne, Germany
dagmar.kern@gesis.org

Peter Mutschke
GESIS – Leibniz Institute for the Social Sciences
Unter Sachsenhausen 6-8
50667 Cologne, Germany
peter.mutschke@gesis.org

Philipp Mayr
GESIS – Leibniz Institute for the Social Sciences
Unter Sachsenhausen 6-8
50667 Cologne, Germany
philipp.mayr@gesis.org

## ABSTRACT
We propose an online access panel to support the evaluation process of Interactive Information Retrieval (IIR) systems - called IIRpanel. By maintaining an online access panel with users of IIR systems we assume that the recurring effort to recruit participants for web-based as well as for lab studies can be minimized. We target on using the online access panel not only for our own development processes but to open it for other interested researchers in the field of IIR. In this paper we present the concept of IIRpanel as well as first implementation details.

## Categories and Subject Descriptors
H.3.2 [**Information Interfaces and Presentation**]: User Interfaces – Evaluation/methodology, User-centered design

## 1. INTRODUCTION
Interactive Information Retrieval (IIR) becomes more and more important in IR research and thus, user involvement in the developing process of IR systems is essential to produce useful and usable interactive products [3]. Following a user-centered design approach [5] the development of different IIR systems includes user involvement in every step of the design cycle through contextual observations, web-surveys, usability tests, quantitative ratings, retrieval tests and performance measures. However, one of the probably most time-consuming issue during the evaluation process of new IIR systems is recruiting participants for online-surveys, interviews or lab studies. Tools like Mechanical Turk[1] are commonly used for performing short web-based user studies [4]. In a relative short period of time a large number of participants can be reached. Mechanical Turk is well suitable for user studies targeting on common internet users with no specific educational or contextual background. Addressing a specific user group with special skills or informational settings is however not reliably possible, and therefore its benefit for evaluating specific IIR systems and tasks seems to be very low.

Online access panels, in contrary, provide the opportunity to limit a sample of participants to a specific target group. While the usage of such a panel is common for market research[2], for data collection in the Social Sciences[3] or even for evaluating existing websites[4] it is often not considered during web product development processes, and hence have – to the best of our knowledge – so far not been taken into account for the continuous development of IIR systems.

An online access panel is a pool of previously recruited participants who have agreed to take part in regularly (web-based) studies [2]. Online access panels with panelists representing the desired target group or with even real end-users of existing products could help to speed-up the recruiting process for user studies and the management of participants who are needed for repeated examinations. A further major benefit of an online access panel is seen in the opportunity to continuously investigate novel IIR approaches with a stable population of users. Follow-up studies with the same user groups might also provide information about changes of information needs over time. But there are also disadvantages like the problem of representativeness (this will be discussed later on in this paper) or the effort for establishing and maintaining such an online access panel. Assuming that the IIR community has the same or at least a similar target group in mind while developing new IIR functionalities, we propose to set up an online access panel for IIR research which is developed and maintained in cooperation. We target on establishing an international group of researchers who collaborate in building up this unique instrument for sharing panelists in one dedicated place.

In the following we describe the basic concept of the proposed online access panel – called IIRpanel[5], its architecture and its initial implementation at GESIS.

---

[1] https://www.mturk.com/mturk/welcome

[2] E.g. http://www.usamp.com/panel.html

[3] E.g. http://www.gesis.org/en/services/data-collection/gesis-panel/

[4] E.g. http://panel.webeffective.keynote.com/Default.asp?

[5] www.gesis.org/iirpanel

## 2. IIRpanel – CONCEPT

The initial idea of maintaining an online access panel originates from the need to have a commonly used proband management system for GESIS. So far for nearly all user studies a new process for recruiting participants has been started. The online access panel in contrast aims at establishing a pool of real users of IIR systems having a strong interest in taking part either in improvement processes of existing systems or in evaluating new IIR functionalities. The idea is that participants sign up only once for the IIRpanel and researchers performing user studies can send participant calls to all or to a sample of this panel with very little effort.

On a sign-up page the users are asked for demographical data like gender, year of birth, their current job, their current profession, research field of interest, and so on. The users can decide if they would like to take part in web-surveys only, in lab studies only or both. For the lab studies we also asked for a location to better arrange lab studies with participants available in a specific area. All this data is stored in a database taking current law of privacy into account. Through a self-service page the user has always access to his data, the ability to delete his own profile, to change his demographical data, to communicate with the panel provider and to access his participation history or even to watch results of studies in which he took part.

Collaborating internal and external researchers performing a user study can log into the IIRpanel and select a sample based on the stored data. An email with a web-survey link or an invitation for a lab user study can be send directly through the IIRpanel. After the user study is finished further administrative data should be stored like title of the study, date of invitation, if the user participated in the study and the date of the study. The participant history which is derived from this information easily allows the researchers to perform a follow-up study with the same set of users.

The initial setup of the panel proposed does not aim on satisfying the claim of representativeness, an often criticized aspect of open access panels [6]. Our main intention is to establish a set of volunteered probands in a specific research domain who are willing to take part in a series of web surveys as well as in lab studies. For now we presume that offering new user-tailored IIR functionalities that make participants' future work easier will be incentive enough for their participation in web-surveys. As lab studies go along with more effort participants will be remunerated for it.

## 3. IIRpanel – ARCHITECTURE

As a basis for the IIRpanel we used phpPanelAdmin developed by Anja Göritz [2]. phpAdmin[6] is an open source tool under the GNU license for managing online access panels or similar sampling lists. As a web-based platform phpPanelAdmin aims at quickly setting up and managing online panels. As the name already suggests it is developed in php and makes use of HTML and JavaScript. For panelists' data storage a MySQL database is required. The administrative functionalities include searching for panelists, managing panelists' data and variables, identifying duplicates, drawing samples, creating and managing e-mail templates, sending e-mails to panelists, and displaying panel statistics [2]. To fulfill our requirements we completed the tool with additional functionalities for user self-services as well as for managing participation history. In Figure 1 the registration form of the current implementation of the online access panel is shown.

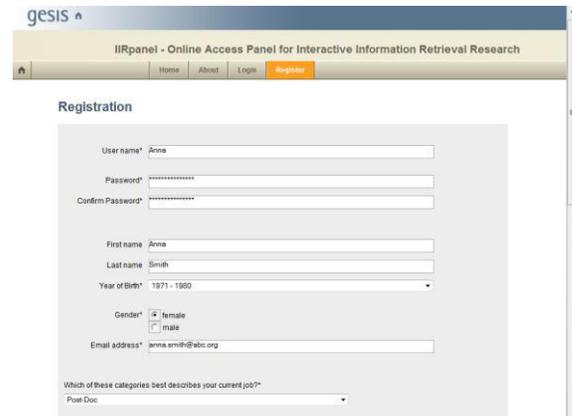

**Figure 1. IIRpanel - Registration form**

## 4. CONCLUSION AND OUTLOOK

We introduced a collaborative online access panel for supporting evaluations of IIR systems. The panel infrastructure proposed aims at minimizing the recruiting process of participants for web-surveys as well as for lab studies and at building-up an international pool of various types of IIR system users to speed up IIR evaluations in the future. Our next steps are recruiting panelists through announcements on our websites as well as in our IR systems, by telephone calls, mailings and word-of-mouth advertising (according to [1]), evaluating a pilot application of the panel with GESIS users, and finally establishing a consortium of international researchers who are interested in supporting the maintenance as well as the further development of the IIR panel. By using IIRpanel we expect benefits for major IIR research issues, such as defining and formulation of adequate search tasks and queries, information behavior, needs over time, usability of search interfaces, utility and usability of search histories, evaluation of whole search sessions performing longitudinal studies on the evolution of information needs, but also the benefits and drawbacks of such an online access panel itself can be addressed as a new area in IIR research.

---

[6] www.goeritz.net/panelware